\documentclass[a4paper, 10pt, conference]{ieeeconf}
\IEEEoverridecommandlockouts
\usepackage{cite}
\usepackage{amsmath,amssymb,amsfonts}
\usepackage{algorithmic}
\usepackage{graphicx}
\usepackage{pdfpages}
\usepackage{textcomp}
\usepackage{xcolor}
\usepackage{multirow}
\usepackage{pifont}
\usepackage{color}
\usepackage{alltt}
\usepackage[hidelinks]{hyperref}
\usepackage{enumerate}
\usepackage{siunitx}
\usepackage{breakurl}
\usepackage{epstopdf}
\usepackage{pbox}
\usepackage{subcaption}
\usepackage{multicol}
\usepackage{float}
\def\BibTeX{{\rm B\kern-.05em{\sc i\kern-.025em b}\kern-.08em
    T\kern-.1667em\lower.7ex\hbox{E}\kern-.125emX}}
\title{\LARGE \bf
    Hybrid Position/Force Control for Hydraulic Actuators
}

\author{Philipp Pasolli and Michael Ruderman% <-this % stops a space
    \thanks{This work has received funding from the EUs
        H2020-MSCA-RISE research and innovation programme under grant
        agreement No 734832.}% <-this % stops a space
    \thanks{P Pasolli and M Ruderman are with University of Agder, Norway \newline
        {\tt \footnotesize \{philipp.pasolli,michael.ruderman\}@uia.no}}%
    \thanks{\textcolor[rgb]{0.00,0.00,1.00}{Accepted article (MED'2020 conference). \textcopyright 2020 IEEE.
    Personal use of this material is permitted. Permission from IEEE must be
    obtained for all other uses, in any current or future media, including
    reprinting/republishing this material for advertising or promotional purposes,
    creating new collective works, for resale or redistribution to servers or lists,
    or reuse of any copyrighted component of this work in other works.}}
}

\begin{document}

    \maketitle
    \thispagestyle{empty}
    \pagestyle{empty}

\begin{abstract}
    In this paper a novel hybrid position/force control with
    autonomous switching between both control modes is introduced for
    hydraulic actuators. A hybrid position/force control structure with
    feed-forwarding, full-state feedback, including integral control
    error, pre-compensator of the dead-zone, and low-pass
    filtering of the control value is designed. Controller gains are
    obtained via local linearization and pole placement accomplished
    separately for the position and force control. A hysteresis-based
    autonomous switching is integrated into the closed control loop,
    while multiple Lyapunov function based approach is applied for
    stability analysis of the entire hybrid control system.
    Experimental evaluation is shown on the developed test setup with
    the standard industrial hydraulic cylinders, and that for
    different motion and load profiles.
\end{abstract}

%\begin{IEEEkeywords}
%Force control, hybrid dynamics, hydraulic
%actuator, motion control, nonlinear systems
%\end{IEEEkeywords}

\bstctlcite{references:BSTcontrol}

\section{Introduction}
\label{sec:I}

Different actuators are in use in the mechatronic applications.
While electric motors, linear drives, or pneumatic actuators are
well suitable for a fast system response, hydraulic actuators are
still the first choice if compact form factor combined with high
power density and reliability are demanded, see
\cite{Jelali2003,Merritt1967} for backgrounds. At the same time,
hydraulic systems are also well known for their nonlinearities and
associated challenges arising during operation in the closed-loop
force or motion control.

Several applications using hydraulic actuators are repetitive and
tedious for human operators, for example excavators, but remain
yet to be controlled manually. While an, at least, semi-automatic
control emerges in these fields, the widespread PID-controllers
keep yet standard also in those applications. While improvements
like for instance optimal design of PID-control with non-linear
extensions were reported \cite{Liu2000}, other research promotes
different control strategies like for example adaptive control or
variable structure control, showing often a superior performance
to the classic PID-controllers, see e.g.
\cite{Koch2016,Vazquez2014,BinYao2000}. When a hydraulically
actuated equipment or machine interacts with the environment, like
given example of an excavator, a permanent use of position control
can become even destructive due to inherently stiff control
properties. In that case a force-based control approach is
necessary. Several force control strategies for hydraulic systems
were reported \cite{Alleyne2000,Komsta2013,Niksefat2000}, while
the force control issues are equally well-known in robotics and
mechatronics, see e.g. \cite{Katsura2006}. However, for a fully
automated operation of such equipment, a combination of force and
position control should be taken into consideration, where the
individual controllers can be switched, correspondingly
reconfigured upon the motion constraints and interaction with the
environment. Such hybrid control approaches remain topical, even
though addressed generally in the former works, especially in the
field of robotics, see e.g. \cite{Raibert1981,Khatib1987}.

In this paper, a hybrid position and force control approach is
pursued based on the feed-forwarding and full-state feedback,
including the integral control error. An appropriate
hysteresis-based switching strategy is integrated into the closed
feedback loop, while changing between the parameter settings of
both control modes. Remarkable is that the derived control
structure self does not change and, therefore, represents an
uniform architecture for the position and force control
simultaneously.
%Some preliminary results of this work have been
%partially reported in \cite{Pasolli2018,Pasolli2019}.

The synthesized hybrid control relies on a detailed model of the
system plant under consideration. In \cite{Ruderman2017a}, a
reduced hydraulic model related to the setup was derived which was
expanded upon in \cite{Pasolli2018}, and the initial hybrid
position-force control results reported in \cite{Pasolli2019}. In
the recent work, we present the fully elaborated hybrid
position/force control with extended design, analysis, and
evaluation. The rest of the paper is structured as following. In
Section II the model from \cite{Pasolli2018} is summarized while
in Section III the proposed control architecture, control gains
tuning, linearized system behavior, and local stability analysis
based on \cite{Ruderman2019} are described. An exhaustive
experimental control evaluation is reported in Section IV with the
paper's summary given in Section V.

\section{System modeling}
\label{sec:II}

The modeled hydraulic system is a single rod, double acting
cylinder connected to a servo valve which in return is connected
to an HPU (Hydraulic Power Unit). The full model from
\cite{Pasolli2018} describes the second order valve dynamics,
dead-zone-saturation combination, orifice and continuity
equations, cylinder dynamics and Stribeck friction. The respective
parameters and characteristics were identified and the simulation
results validated by comparing with experimental data from
different tests. A model reduction was performed under a mild
assumption of equal cylinder cross sections, thus introducing a
total load dependent pressure and flow. The control valve dynamics
were neglected due to an observed unity gain in the frequency
range of interest. Linearisation were performed of the form $g(x)
\approx kx+d$, with $k$ being the slope and $d$ the offset of a
linear function, whose nonlinear counterpart is $g(\cdot)$. The
linearized, piecewise affine, state-space model of the plant is
then given by
\begin{equation}
\label{eq:OL_state_space}
\begin{aligned}
\dot{\textbf{x}}&=\textbf{A}\textbf{x}+\textbf{b}u + \textbf{f},\\
y &= \textbf{c}\textbf{x},
\end{aligned}
\end{equation}
where $\textbf{x}= (x,\dot{x},P_L,F_L)^T$ is the state vector with
$x$ being the cylinder position, $\dot{x}$ the cylinder velocity,
$P_L$ the load dependent pressure and $F_L$ the external load
force. The system matrix, input coupling, affine term, and output
coupling vectors are given by
\begin{equation}
\label{eq:ol_system_matrix}
\begin{aligned}
\textbf{A} &=
\begin{bmatrix}
0&1&0&0 \\[3pt]
0&-\begin{aligned}\frac{k_w}{m}\end{aligned}&\begin{aligned}\frac{\bar{A}}{m}\end{aligned}&- \begin{aligned}\frac{1}{m}\end{aligned} \\[3pt]
\begin{aligned} 0 \end{aligned}& \begin{aligned} -\frac{4E\bar{A}}{V_t} \end{aligned}  & \begin{aligned} \frac{4E\hat{C}_{qp}}{V_t}\end{aligned} & \begin{aligned} 0\end{aligned}  \\[3pt]
0&0&0&0\\
\end{bmatrix},
\end{aligned}
\end{equation}
\begin{equation}
\begin{split}
\textbf{b}=
\begin{bmatrix}
\label{eq:ol_system_vector}
0\\[0pt]
0\\[0pt]
\frac{4E\hat{C}_qk_g}{V_t}\\[0pt]
0
\end{bmatrix},
\end{split}
\quad
\begin{split}
\textbf{f} =
\begin{bmatrix}
0\\[3pt]
-\frac{d_w}{m}\\[0pt]
\frac{4E\hat{C}_qd_g}{V_t}\\[0pt]
0
\end{bmatrix},
\end{split}
\quad
\begin{split}
\textbf{c}^T =
\begin{bmatrix}
1\\[0pt]
0\\[0pt]
0\\[0pt]
0
\end{bmatrix},
\end{split}
\end{equation}
correspondingly. Here $k_w$ and $d_w$ are coefficients of the
linearized Stribeck function, $m$ is the lumped moving mass of the
cylinder, $\bar{A}$ is the average cylinder cross section, $E$ is
the hydraulic fluids bulk modulus, $V_t$ is the combined hydraulic
fluids volume in the lines from the valve to the cylinder,
$\hat{C}_q$ and $\hat{C}_{qp}$ are the coefficients of the
linearized orifice equation and $k_g$ and $d_g$ are the
coefficients of the linearized dead-zone and saturation
combination. For more model details we refer to
\cite{Pasolli2018}. It is worth emphasizing that the modeled load
force, as a state, provides zero eigen-dynamics since being an
exogenous external quantity. Note that the above state-space model
is configured here to have the piston position as the output
value. This will be reconfigured when switching between the
position and force control.

\section{Control Design}
\label{sec:III}

This section is dedicated to control design, while describing the
overall hybrid closed-loop system. The determination of control
parameters and local stability are also addressed. For details on the
residual control components, such as feed-forward dead-zone
compensation, signal filtering, and switching strategy we refer to
\cite{Pasolli2019} due to the space limitations.

\subsection{Hybrid control structure}
\label{sec:control_structure}

The proposed control architecture includes the feed-forward,
integral-error- and full-state-feedback. A more classical
architecture would refer to a cascaded structure, where the
inner-loop represents a force control and the outer-loop the
position control. Such control approaches have certain inherent
shortcoming. An ideal force control tends to zero control
stiffness, while an ideal position control stiffness tends towards
infinity. Therefore a cascaded combination would always constitute
a tradeoff and offer a suboptimal performance when targeting the
position and force control tasks simultaneously.

\begin{figure}[!h]
    \centering
    \includegraphics[width=0.9\linewidth]{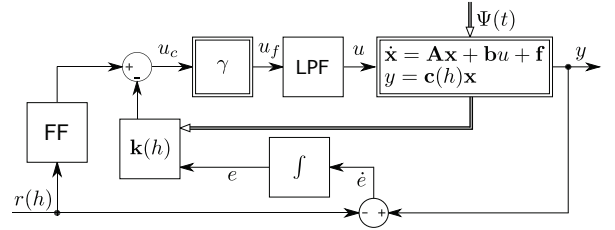}
    \caption{Block-diagram of the hybrid position/force control}
    \label{fig:controlstructure}
\end{figure}

For enhancing the control performance of both operation modes, a
hybrid switched position/force control is developed, see Fig.
\ref{fig:controlstructure}. Note that an additional vector of
external disturbances $\Psi(t)$ is drawn, for the sake of
completeness, although not explicitly modeled and analyzed, that
due to a limited process knowledge. For the rest of the paper, $h$
represents the discrete switching variable and, therefore, the
corresponding operation mode, with $h=-1$ for position and
$h=1$ for force control. Note that the hybrid position/force
control structure remains the same upon switching and allows for
control parameters to be determined separately, so as to meet the
performance requirements in both cases.

For the closed-loop control system the state vector of
\eqref{eq:OL_state_space} is extended by an additional integral
control error state with $\dot{e}=y - r(h)$ dynamics, thus
resulting in $\textbf{x}_e = (\textbf{x}^T, e)^T$. Here $r(h)$ is
the control reference value, while the output $y$ depends on the
control mode and is switched through the output coupling vector
$\mathbf{c}(h)$. Note that in order to accommodate the integral
error state, the system matrix \eqref{eq:ol_system_matrix} and
vectors \eqref{eq:ol_system_vector} of the state-space model are
directly extendable to $\mathbf{A}_e \in \mathbb{R}^{5 \times 5}$
and $\mathbf{b}_e, \mathbf{c}_e^T, \mathbf{f}_e \in
\mathbb{R}^{5 \times 1}$. The resulting control law is then
\begin{equation}
u_c = \text{FF} \, r(h) - \textbf{k}(h)\mathbf{x}_e,
\end{equation}
where $\mathbf{k}(h) = (K_1, K_2, K_3, K_4, K_i)$ is the vector of
the control gains, which are determined separately for position
and force control. FF represents the feed-forward control part,
which is the single gain value for the given structure, and that
FF $= K_1$ for position and FF $=K_4$ for force control.
The overall control structure, cf. Fig.
\ref{fig:controlstructure}, includes also the pre-filtering
dead-zone compensator $\gamma$ and the low-pass filtering (LPF) of
the control signal, both addressed in \cite{Pasolli2019}.

\subsection{Determining of control gains}
\label{subsec:IIIe}

The control gain parameters are determined via standard pole
placement accomplished for the linearized models, that separately
for the position and force control modes. The pole placement is
made for the test case scenarios, the same which are lately
evaluated in the experiments. The controlled rod displacement is
driven with a constant speed until it reaches a hard stop (by
environment), that triggers an autonomous switching to the force
control at which the rod is holding a constant force. Since the
steady-state velocity and force are defined by reference, the
residual state values required for the model linearization are
extrapolated from the simulation of the full-order (nonlinear)
model, cf. \cite{Pasolli2018}.

The state-space model \eqref{eq:OL_state_space} does not allow for
directly using the pole placement, that due to inclusion of the
affine terms. In order to deal with affine vector $\mathbf{f}_e$,
that when deriving the state-space form applicable for a pole
placement, the state vector is further extended to
$\bar{\mathbf{x}} = [\mathbf{x}_e^T, 1]^T$, thus resulting in
\begin{equation}
\dot{\bar{\textbf{x}}} = \begin{bmatrix} \dot{\mathbf{x}}_e \\
0\end{bmatrix} = \bar{\mathbf{A}}\bar{\mathbf{x}} +
\bar{\mathbf{b}} \, r =
\begin{bmatrix}\mathbf{A}_e & \mathbf{f}_e \\
\mathbf{0} & 0
\end{bmatrix} \bar{\mathbf{x}} + \begin{bmatrix} \mathbf{b}_e \\ 0
\end{bmatrix} r,
\end{equation}
while the output coupling vector is extended to $\bar{\mathbf{c}}
= [\mathbf{c}_e,0]$. This allows establishing the set of two
state-space forms, where the switching discrete state
$h=\left[-1,1\right]$ refers to the position and force control
respectively. The overall hybrid control system, in the linearized
form, is then given by
\begin{equation}
\label{eq:linearized}
\begin{aligned}
\dot{\bar{\textbf{x}}}&= \bar{\textbf{A}}(h)\bar{\textbf{x}}+\bar{\textbf{b}}(h)r(h),\\
y &= \bar{\textbf{c}}(h)\bar{\textbf{x}}.
\end{aligned}
\end{equation}

Since the feedback control design relies on the linearized
modeling at steady-state operational conditions, following
assumptions, correspondingly simplifications, can be made for
obtaining the constant system matrices and vectors of
\eqref{eq:linearized}. For the steady-state velocity, equally as
steady-state reactive force of the environment, a steady-state
load pressure of the cylinder can be assumed. This is inherent
since the load pressure is equivalent to the hydraulic driving
force, correspondingly pressure difference between both chambers.
Therefore, the orifice equation, cf. \cite{Pasolli2018}, can be
simplified to
\begin{equation}
\bar{Q}_L = z K \Omega,
\end{equation}
where $\Omega= \sqrt{0.5(P_S - P_L)}$ for $z>0$, and it is valid
$0 \leq \Omega \leq \sqrt{0.5 P_S} $ for $|P_L| = \mathrm{const} <
P_S$ at the steady-state. $K$ is the valves' flow coefficient and
$z$ refers to the orifice opening of the valve, including the
combination of dead-zone and saturation. One can define $z =: u_c$
with saturations, that due to cancelation of the dead-zone by the
forward compensator $\gamma$. Since the numerical simulation does
not highlight saturated control values, that for the reference
scenarios under evaluation, the nonlinear saturation by-effect can
also be neglected when tuning the linear control parameters. Also the
dynamic behavior of LPF, cf. with Fig. \ref{fig:controlstructure},
is neglected since the LPF bandwidth was set to coincide with that
of the servo-valve, cf. \cite{Pasolli2019}. Both corner
frequencies (of LPF and servo-valve) are significantly higher compared to dynamics of the
operated hydraulic actuator. With respect to the above
assumptions, the matrices and vectors of \eqref{eq:linearized} are
given by
\begin{equation}
\begin{aligned}
\bar{\textbf{A}}(-1) &=
\begin{bmatrix}
0&1&0&0&0&0 \\[0pt]
0&\begin{aligned}-\frac{k_w}{m}\end{aligned}&\begin{aligned}\frac{\bar{A}}{m}\end{aligned}& \begin{aligned}0\end{aligned}&0&\begin{aligned}-\frac{d_w}{m}\end{aligned} \\[0pt]
\begin{aligned} a_{31} \end{aligned}& \begin{aligned} a_{32} \end{aligned}  & \begin{aligned} a_{33}\end{aligned} & \begin{aligned} a_{34} \end{aligned}   & \begin{aligned} a_{35} \end{aligned}&0\\[0pt]
0&0&0&0&0&0\\[0pt]
-1&0&0&0&0&0\\[0pt]
0&0&0&0&0&0\\[0pt]
\end{bmatrix},
\end{aligned}
\end{equation}
\begin{equation}
\label{eq:b_pos}
\begin{aligned}
\bar{\textbf{b}}(-1)=\begin{bmatrix}
0&0&\frac{4E}{V_t}K\Omega K_1&0&1&0
\end{bmatrix}^T,
\end{aligned}
\end{equation}
\begin{equation}
\begin{aligned}
\bar{\textbf{c}}(-1)= \begin{bmatrix} 1&0&0&0&0&0
\end{bmatrix}
\end{aligned}
\end{equation}
for the position control, i.e. $h=-1$, and by
\begin{equation}
\begin{aligned}
\bar{\textbf{A}}(1) &=
\begin{bmatrix}
0&1&0&0&0&0 \\[0pt]
0&\begin{aligned}-\frac{k_w}{m}\end{aligned}&\begin{aligned}\frac{\bar{A}}{m}\end{aligned}& \begin{aligned}-\frac{1}{m}\end{aligned}&0&\begin{aligned}-\frac{d_w}{m}\end{aligned} \\[0pt]
\begin{aligned} a_{31} \end{aligned}& \begin{aligned} a_{32} \end{aligned}  & \begin{aligned} a_{33}\end{aligned} & \begin{aligned} a_{34} \end{aligned}   & \begin{aligned} a_{35} \end{aligned}&0\\[0pt]
0&c&0&0&0&0\\[0pt]
0&0&0&-1&0&0\\[0pt]
0&0&0&0&0&0\\[0pt]
\end{bmatrix},
\end{aligned}
\end{equation}
\begin{equation}
\label{eq:b_force}
\begin{aligned}
\bar{\textbf{b}}(1)=\begin{bmatrix} 0&0&\frac{4E}{V_t}K\Omega
K_4&0&1&0
\end{bmatrix}^T,
\end{aligned}
\end{equation}
\begin{equation}
\begin{aligned}
\bar{\textbf{c}}(1)= \begin{bmatrix} 0&0&0&1&0&0
\end{bmatrix}
\end{aligned}
\end{equation}
for the force control, i.e. $h=1$, respectively. The individual,
summarized for convenience, matrix elements are
\begin{equation}
\label{eq:parameters}
\begin{aligned}
a_{31} &= -\frac{4E}{V_t}K\Omega K_1,\\
a_{32} &= -\frac{4E}{V_t}(K\Omega K_2+\bar{A}),\\
a_{33} &= -\frac{4E}{V_t}K\Omega K_3,\\
a_{34} &= -\frac{4E}{V_t}K\Omega K_4,\\
a_{35} &= \frac{4E}{V_t}K\Omega K_i.
\end{aligned}
\end{equation}
The available nominal system parameters are $\bar{A} = 0.001m^2$,
$m = 1.7026kg$, $K = 0.252e^{-6}\frac{m^3}{s\sqrt{Pa}}$,
$E=10^9Pa$, $V_t = 0.0014m^3$ and $c=1.2e^8N/m$. The latter, which
is an equivalent environmental stiffness, is determined as a
lumped parameter of the coupled rods, force sensor, and material
properties of the cylinder cap. The residual constants are
dependent on which operation mode is active. According to the test
scenarios, $\Omega = 2.23e^3$, $k_w = 1.0151e^3$ and $d_w=30.755$
for the position control, while $\Omega = 2.0125e^3$, $k_w =
6.2499e^3$ and $d_w=0$ for the force control. Recall that $k_w$
represents the slope and $d_w$ the offset of the linearized
Stribeck function. The gain parameters, entering eqs.
\eqref{eq:b_pos}, \eqref{eq:b_force} and \eqref{eq:parameters},
are determined by the pole placement.

\begin{figure}
    \centering
    \includegraphics[width=0.95\linewidth]{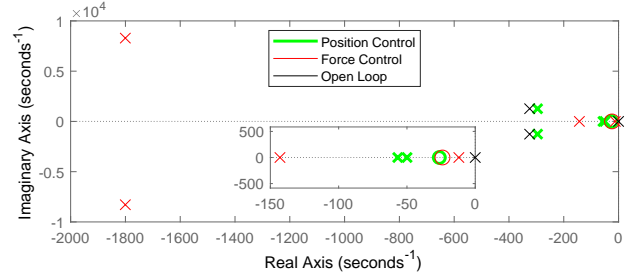}
    \caption{Pole-zero diagram of closed-loop controls versus plant}
    \label{fig:pzmap}
\end{figure}

In Fig. \ref{fig:pzmap}, the poles of both closed-loop controls
are shown versus those of the system plant (open loop). Note that
the most left complex pole pair is associated with hydraulics
behavior, which is marginally affected by adjusting the respective
control gains.
\begin{figure}[!h]
    \centering
    \includegraphics[width=0.95\linewidth]{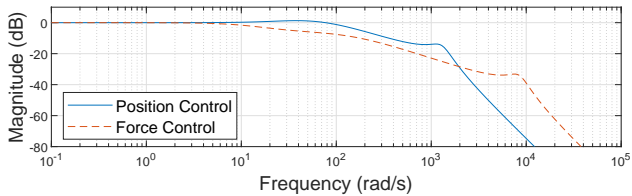}
    \caption{Bode diagram of designed closed-loop controls}
    \label{fig:bode_mag}
\end{figure}
The dominant poles, i.e. closer to origin, refer to the controlled
actuator system dynamics. The gains were chosen so that no complex
pole pairs occur close to the origin, and the most right real pole
satisfies the requirements on the control system dynamics. In
addition, higher gain values were avoided by taking into account
the control saturations and measurement noise included into
numerical simulation.
%While for the simulated closed loop model, faster poles could have
%been chosen, an experimental evaluation showed that faster poles
%result in undesirable behavior like high frequency oscillations of
%the servo valves' spool due to the amplification of the measured
%signals noise and oscillations around the references due to
%un-modeled dynamics like e.g. hydraulic fluid, air in the
%hydraulic lines, etc.
From the Bode plot shown in Fig. \ref{fig:bode_mag}, it can be
seen that the first corner frequency, correspondingly bandwidth,
is about 21Hz for the position control and about 2.5Hz for the
force control. The determined control gains are listed in Table
\ref{tab:gains}.
\begin{table}
    \caption{Control gain parameters}
    \label{tab:gains}
    \begin{center}
        \begin{tabular}{|c|c|c|}
            \hline
            & \textbf{Position Control} & \textbf{Force Control}  \\
            \hline
            $K_1$& 190 & $0$ \\
            \hline
            $K_2$& $9.019e^{-4}$ & $2.5e^{-4}$ \\
            \hline
            $K_3$& $30.539e^{-9}$ & $5.9e^{-8}$ \\
            \hline
            $K_4$& $0$ & $5e^{-5}$ \\
            \hline
            $K_i$& $5000$ & $1.2e^{-3}$ \\
            \hline
        \end{tabular}
    \end{center}
\end{table}
Note that $K_4$ is set to zero for the position
control and $K_1$ is set to zero for the force control, since these states are irrelevant for the respective control mode.

\subsection{Stability analysis}
\label{subsec:sability}
Generally it is desirable to proof a system to be globally
stable. For physical systems that is in most cases however not feasible due the physical restrictions of the experimental setup
which in this case are e.g. maximum pressure, range of motion
of piston stroke, etc. Instead it can be pursued to proof the
stability of the reachable state space. In this paper the proof
of stability is limited to a subset of the reachable state space
proofing local stability, specific for the test scenarios derived,
due to spacial limitations.
The local stability analysis relies on the stability of both
linearized closed-loop control systems, see poles configuration in
Section \ref{subsec:IIIe}, and multiple Lyapunov function approach
applicable to the switched systems, see \cite{Liberzon1999a}. Note
that the stability of switching between the motion and force
control has been recently discussed in detail in
\cite{Ruderman2019}, that for the linearized closed-loop behavior
and autonomous switching by means of a hysteresis relay. Thus, we
give here the main statement only, while for details on using the
multiple Lyapunov function an interested reader is referred to
\cite{Liberzon1999a} in general and to \cite{Ruderman2019} for the
motion and force control systems more specifically. For both
closed-loop control systems, given by \eqref{eq:linearized}, the
quadratic Lyapunov function candidate can be assumed as
\begin{equation}
\label{eq:lyapunov} L(h) = W_1 \dot{x}^2 + W_2 P_L^2 + W_3
\bigl(r(h) - y(h)\bigr)^2.
\end{equation}
Note that this contains all terms related to an energy storage in
the control system: kinetic energy of the relative motion,
potential energy of hydraulic pressure and potential energy of the
feedback control loop reflected through the quadratic control
error. The positive coefficients $W_{1-3}$ can be found for
Lyapunov stability proof and used as mode-dependent, i.e.
$W_i(h)$, that for the sake of better visualization/comparison of
the multiple Lyapunov function.
%\begin{figure}[]
%    \centering
%    \includegraphics[width=1\linewidth]{figures/MultLyapun}
%    \caption{Multiple Lyapunov function for position (red solid line) and force (blue dashed line) controls at periodic switching}
%    \label{fig:lyapunov}
%\end{figure}
%The multiple Lyapunov function of both control modes is shown in
%Fig. \ref{fig:lyapunov} for a periodic switching, meaning an
%alternating position and force control mode.
A non-increase in the Lyapunov function level for two consecutive
operations of the same mode implies the entire switched system is
asymptotically stable, cf. \cite{Ruderman2019}, in this case meaning an alternating position and force
control mode. Since both closed-loop dynamics are linearized and
stable in vicinity to the switching point, the above local
stability of the switching between the modes appears sufficient
for analysis of the designed hybrid control system and its test scenarios.

\section{Experimental Evaluation}
\label{sec:IV}

\begin{figure}[!h]
    \centering
    \includegraphics[width=0.95\linewidth]{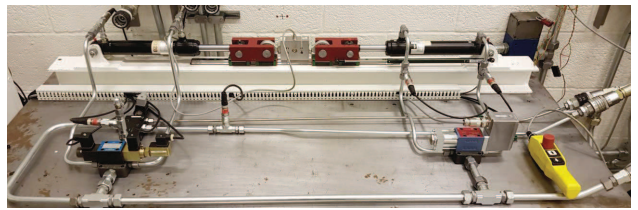}
    \caption{Experimental setup of hydraulic actuators}
    \label{fig:experimental-setup}
\end{figure}

An experimental evaluation of the proposed hybrid control is given
below. The laboratory view of the experimental setup is shown in
Fig. \ref{fig:experimental-setup} while for more details the
reader is referred to \cite{Pasolli2018}.
Two test scenarios are considered: in the first one the controlled
motion of the piston rod of the right-hand cylinder is executed
until hard-stop at the cylinder boundary. In the second one a slow
dynamic counteracting force is introduced during which the piston
rod is moving within $0 \leq x < l$, since
a hydraulic actuator is unsuitable for high
dynamic force control applications \cite{Alleyne1999}. The counteracting force
is produced by another (left-hand) cylinder connected to the
bidirectional control valve (BDCV), which changes the flow
direction, and pressure relief/reduction valve (PRV), to adjust pressure. This allows producing an adjustable (open-loop
controlled) force on the stiff interface between both cylinders.
The supply pressure was set to $P_S = 100e^5$ Pa, and
flow to $40$ l/min.

\subsection{Hard-stop environment}

\begin{figure}[!h]
    \centering
    \includegraphics[width=0.95\linewidth]{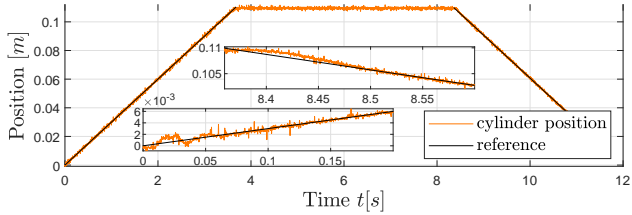}
    \caption{Reference and measured position control response}
    \label{fig:wallpos}
\end{figure}

The controlled motion starts at zero position (controlled
right-hand cylinder fully retracted) and follows the ramp
reference with a slope corresponding to 0.03 m/s velocity, cf.
Fig. \ref{fig:wallpos}. The controlled motion reaches hard-stop
when the overall drive of both cylinders, connected stiffly,
reaches mechanical boundary of the left-hand cylinder. Note that
during the experiment the BDCV is in zero position, thus opening
both chamber lines of the left-hand cylinder to the tank and,
thus, providing no active counteraction force but passive
additional load only. When reaching hard-stop by environment, the
counteracting force rises, due to the stiff position control, and
the hysteresis relay-based switching triggers the force control at
the set threshold value. The force reference trajectory is
initially set to the constant value $r(1) = 3500N$ and afterwards
decreases towards zero in a slow cosine shape. This reference
trajectory is assigned in order to evaluate simultaneously the set
value and trajectory following of the force control and, the
autonomous switching back to the position control. The latter
occurs when passing the lower threshold value of the load force,
that means releasing from the contact with environment. After
switching back, the position control tracks the negative ramp,
with a slope corresponding to $-0.03$m/s velocity, that until
reaching the initial zero position.

Figure \ref{fig:wallpos} shows the reference and measured position
values of the right-hand cylinder rod. The transient phases at the
beginning of relative motion and after switching back (from the
force to position control) are additionally zoomed-in around the
time of 0.1 sec and 8.5 sec correspondingly. Accurate reference
following with solely minor transient overshoots can be
recognized.
\begin{figure}[!h]
    \centering
    \includegraphics[width=0.95\linewidth]{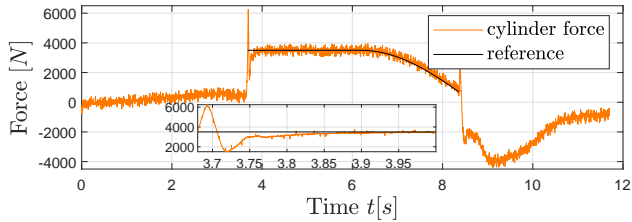}
    \caption{Reference and measured force control response}
    \label{fig:wallforce}
\end{figure}

Figure \ref{fig:wallforce} shows the sensor measured counteraction
force versus the corresponding reference. Note that the triggered
force control is active only during the time span where the
reference is indicated, while the residual force measurements
correspond to both slopes of the controlled motion, i.e. position
control. There, the load force occurs between the right-hand
(driving) cylinder and left-hand (driven) cylinder. An accurate
force following can be recognized, while some transient swinging
occurs at the hard-stop contact, which is an inherent and
well-known issue of the force control in general, compare with e.g.
\cite{Alleyne2000,Katsura2006}.

%The control value after passing through the filter and dead-zone
%compensator is shown in Fig. \ref{fig:wallmoogin}. Note, that the
%control signal shows an oscillatory behavior during force control
%where only the values above 0.1 and below -0.1 actually affect the
%cylinders pressure/motion due the inherent dead-zone of the valve.
%\begin{figure}[H]
%    \centering
%    \includegraphics[width=1\linewidth]{figures/wall_moogin}
%    \caption{Control signal after dead-zone compensator and filter}
%    \label{fig:wallmoogin}
%\end{figure}

\begin{figure}[!h]
    \begin{subfigure}[b]{0.23\textwidth}
        \includegraphics[width=\textwidth]{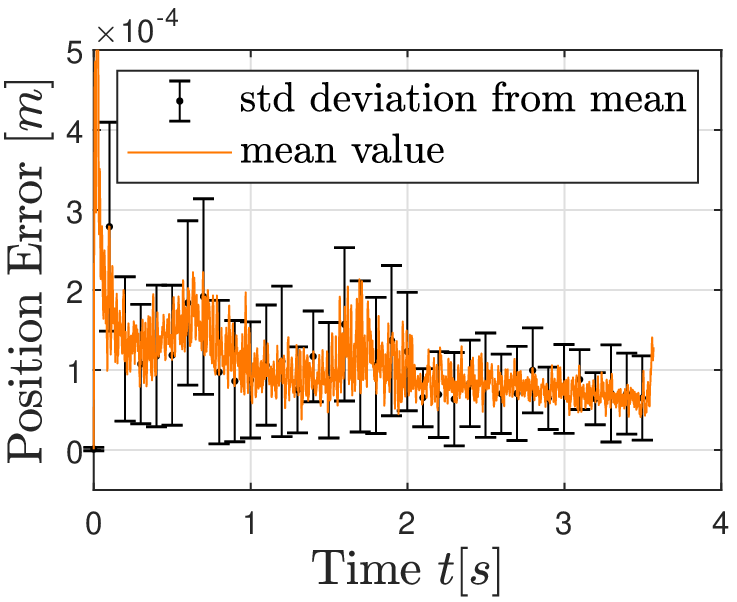}
        \caption{extending}
        \label{fig:wallposerror1}
    \end{subfigure}
    \begin{subfigure}[b]{0.23\textwidth}
        \includegraphics[width=\textwidth]{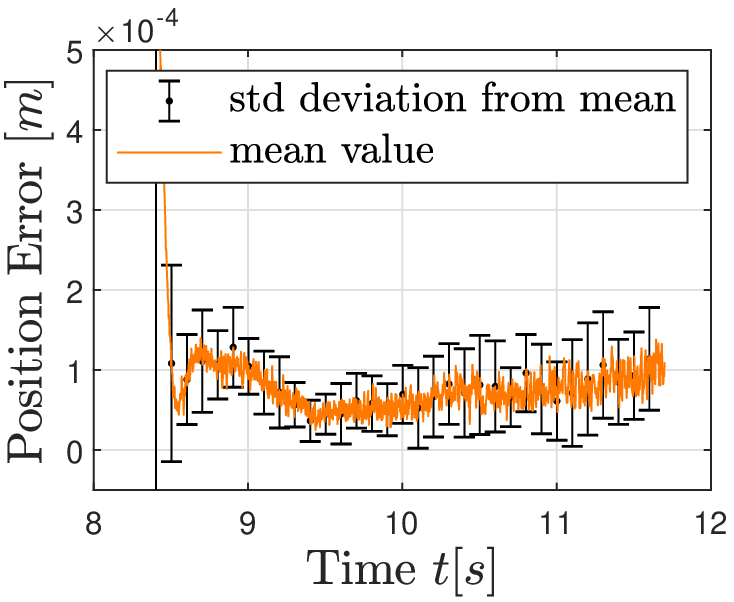}
        \caption{retracting}
        \label{fig:wallposerror2}
    \end{subfigure}
    \caption{Position control error over 20 experiments}
\end{figure}

\begin{figure}[!h]
    \centering
    \includegraphics[width=0.95\linewidth]{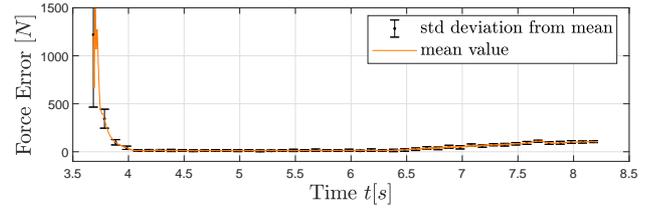}
    \caption{Force control error over 20 experiments}
    \label{fig:wallforceerror}
\end{figure}

In order to assess repeatability of the proposed hybrid control
scheme, a total of 20 experiments were conducted. The mean values
and standard deviations of the control errors are evaluated from
the signals processed by the second-order LPF with 100 Hz cutoff
frequency. The low-pass filtering is done for the sake of better
visualization of all tests against each other, since the inherent
level of process and measurement noise in the hydraulic system is
relatively high. Figures \ref{fig:wallposerror1} and
\ref{fig:wallposerror2} show the absolute mean values and standard
deviations of the position control error for both ramp segments of
the rod displacement. The absolute mean values and standard
deviations of the force control errors are shown in Fig.
\ref{fig:wallforceerror}.

\subsection{Dynamic environment}

\begin{figure}[!h]
    \centering
    \includegraphics[width=0.95\linewidth]{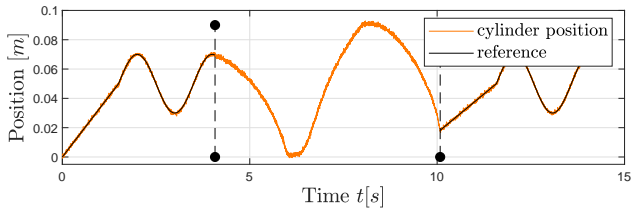}
    \caption{Reference and measured position control response}
    \label{fig:dynpos}
\end{figure}

\begin{figure}[!h]
    \centering
    \includegraphics[width=0.95\linewidth]{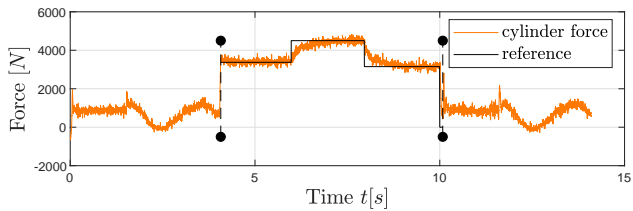}
    \caption{Reference and measured force control response}
    \label{fig:dynforce}
\end{figure}

\begin{figure}[!h]
    \begin{subfigure}[b]{0.23\textwidth}
        \includegraphics[width=\textwidth]{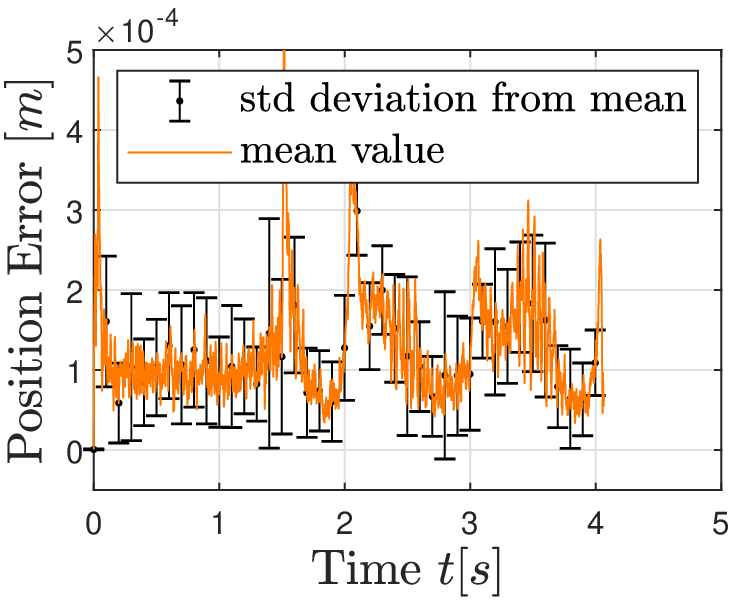}
        \caption{initial}
        \label{fig:dynposerror1}
    \end{subfigure}
    \begin{subfigure}[b]{0.23\textwidth}
        \includegraphics[width=\textwidth]{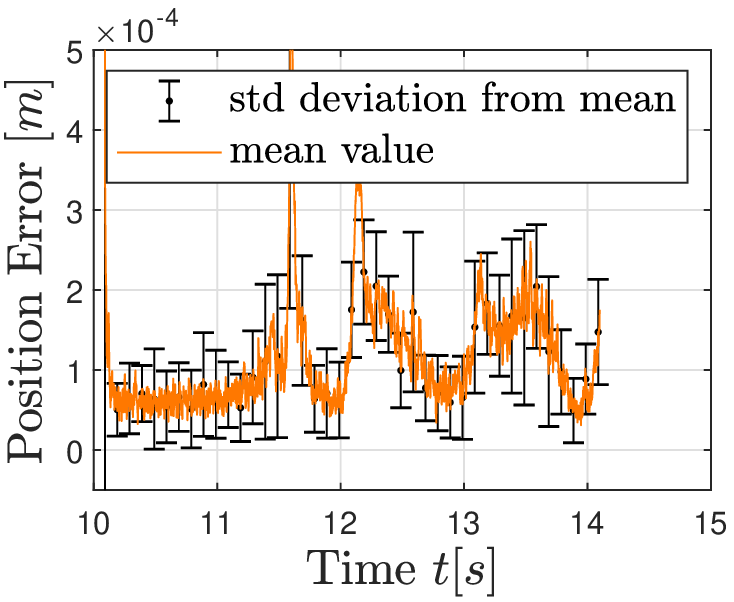}
        \caption{last}
        \label{fig:dynposerror2}
    \end{subfigure}
    \caption{Position control error over 20 experiments}
\end{figure}

\begin{figure}[!h]
    \centering
    \includegraphics[width=0.95\linewidth]{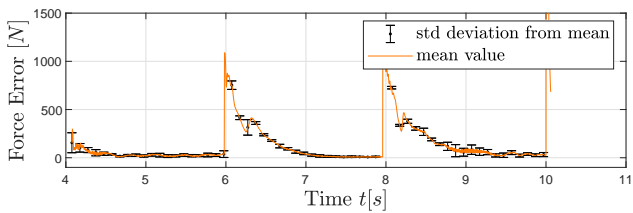}
    \caption{Force control error over 20 experiments}
    \label{fig:dynforceerror}
\end{figure}

For the test scenario with dynamic environment, the controlled
motion starts at fully retracted zero position, following a ramp
reference within first 1.5 sec, after which the reference
trajectory becomes a slow sinusoidal pattern, cf. Fig.
\ref{fig:dynpos}. Note that the reference trajectory $r(-1)$ is
shown for the position control mode only; after switching back
from the force control, the position reference is recalculated on the
fly. The left-hand cylinder is feed-forward controlled in an
open-loop manner, while being stiffly coupled to the right-hand
(controlled) one. This way, a varying counteraction force is
generated. The BDCV control value, for the left-hand cylinder, is
chosen such that it is continuously driving the left-hand
cylinder, while the PRV is controlled in a pulsed pattern with 10
sec period and 60\% pulse-width. The force reference trajectory
$r(1)$ starts at 3375 N and then steps up and down to 4500 N and
3150 N respectively, see Fig. \ref{fig:dynforce}. These step
values are chosen such that the controlled cylinder rod cannot
reach fully extended/retracted states while pushing against the
left-hand cylinder rod.
%The PRV reaching pulse based low-level causes again a mode switch
%from force to position control, at which time a linear reference
%function is calculated on the fly, causing the cylinder to reach a
%position of $x=0.05$m within $1.5$s, after which the previously
%described sinusoidal motion is continued.
Figure \ref{fig:dynpos} shows the position measurement together
with the reference trajectory segments. After 4 sec time, labeled
by the dashed bar, the PRV controlling the left-hand load cylinder
switches to high, thus resulting in an increased load force and,
hence, triggering switch to the force control mode.
The corresponding reference force and (all time) measured load
force can be seen in Fig. \ref{fig:dynforce}. After 10 sec time,
the PRV control value is again at low-level, the load force drops
below the relay switching threshold and the hybrid controller
changes back to the position control mode, labeled by the second
dashed bar in Fig. \ref{fig:dynforce}. Note that between both
dashed bars, a relative motion occurs (see Fig. \ref{fig:dynpos}),
while the required controlled force is kept constant, cf. Fig.
\ref{fig:dynforce}. Also here the experiments were repeated 20
times, while the same signals filtering as described above was
applied when evaluating the absolute mean values and standard
deviation of the control error. Both are shown in Figs.
\ref{fig:dynposerror1} and \ref{fig:dynposerror2} for the position
control mode, and in Fig. \ref{fig:dynforceerror} for force
control mode respectively.

\section{Summary}

A novel hybrid position/force control suitable for standard
linear-stroke hydraulic actuators was proposed. For its
development an experimental hydraulic test rig with two cylinders
in antagonistic setup was designed, assembled, and instrumented.
The corresponding full-order model of the system was derived. A
reduction of the state-space model was made, forming basis for the
hybrid control loop design. The included autonomous switching
between the position and force control relies on the hysteresis
relay, that without changing the overall control structure.
Control parameters were obtained based on the pole placement. The
local stability around the operation and switching points was
shown. Two experimental studies were illustrated for evaluating
the repeatability and performance of both, position and force
controls, and switching between them.

% References
\bibliographystyle{ieeetran}
\bibliography{references}

\end{document}